# The Need, Benefits, and Demonstration of a Minimization Principle for Excited States

Naoum C. Bacalis

*Theoretical and Physical Chemistry Institute, National Hellenic Research Foundation, Vasileos Constantinou 48, GR-116 35 Athens, Greece*

**Abstract.** It is shown that the standard methods of computing excited states in truncated spaces **must** yield wave functions that, beyond truncation, are in principle veered away from the exact, and a remedy is demonstrated via a presented functional, $F_n$, obeying a minimization principle for excited states. It is further demonstrated that near avoided crossings, between two MCSCF "flipped roots" the wave function that leads to the excited state has the lowest $F_n$.

## THE NEED AND BEBEFITS OF A MINIMIZATION PRINCIPLE FOR EXCITED STATES

In this presentation two subjects, constituting the essence of the "truncation problem", and their remedy are addressed. The central idea can be exposed by an example: Consider, for example, He $1s^2$ $^1S$: Due to the electron repulsion the orbital $1s$, doubly occupied, is more diffuse than singly occupied. When excited to $1s2s$ $^1S$, the outer $2s$ electron pushes the inner in, making $1s$ more compact than before. If the previous more diffuse $1s$ is used to describe the excited state $1s2s$, then, in an *ab-initio* calculation, many configurations will be needed to "fix" it more compact, and current methods resort to just huge wave functions to "fix" relatively simple orbitals. But then, large systems cannot be treated. Contrarily, our method finds directly the correct compact $1s$ orbital, independently of the ground state $1s$, and many configurations are not needed: A small space suffices.

If, on the other hand, one *optimizes* the *excited state*, $1s2s$, toward the correct energy, then its (compact) $1s$, used in the *same* truncated space for the description of the orthogonal ground state $1s^2$, deteriorates the ground state, so, an *excited function is obtained orthogonal to a **deteriorated** ground state,* therefore the "optimized" $1s2s$ function, cannot be correct (despite its correct energy, as explained below), veered away from the exact excited function.

Besides, an alternative attempt: i.e. to minimize the energy "*orthogonally to a ground state approximant*" must, as shown below, yield energy not collapsed, nevertheless *below* the exact excited, of course also veered away from the exact excited function. Therefore, in both methods of approach we have a "truncation problem".

Thus, the two issues addressed, and their remedy, are: (i) According to the standard methods for computing excited states, based on the Hylleraas, Undheim, and McDonald (HUM) theorem [1], the eigenvectors in a truncated space of the higher secular equation roots inevitably **must** be veered away from the corresponding exact excited Hamiltonian eigenfunctions while their eigenvalues are *upper* bounds of the exact energy. (ii) Any energy minimization *orthogonally to lower lying* (truncated) *approximants* is not collapsed to lower lying states, but **must** yield a function that is veered away from the exact Hamiltonian excited eigenfunction [2-5], with energy that is a *lower* bound of the exact energy (after passing the *closest to the exact*, which necessarily lies *below* the exact [3 and cf. below]). However, as Shull and Löwdin have shown [6], the excited states can be computed independently.

Our method, a minimization principle for excited states, $F_n[\phi_n, \{\phi_{i<n}\}]$,

$$F_n = F_n[\phi_n, \{\phi_{i<n}\}] \equiv E[\phi_n] + 2\sum_{i<n} \frac{\left(E[\phi_n]\langle\phi_i|\phi_n\rangle - \langle\phi_i|H|\phi_n\rangle\right)^2}{E[\phi_n] - E[\phi_i]} \left(1 - \sum_{i<n}\langle\phi_i|\phi_n\rangle^2\right)^{-1} \quad (1)$$

preliminarily successfully tested for atoms within variational configuration interaction [2-5], overcomes both of the above problems, approaching variationally the exact excited state wave functions $\psi_n$: $\phi_n \to \psi_n$, (i.e., in the above example, finding the correct compact $1s$ of $1s2s$), as local minima of $F_n$, which do not depend crucially on the ground and lower excited state approximants $\phi_i$, (i.e., in the above example, in Eq.1 the diffuse $1s^2$ is used as $\phi_{i=0}$, which may be not so accurate, only reasonable, [3]) and the minima are located *at* the excited states $\psi_n$ of a non degenerate Hamiltonian in a truncated space (i.e. without resorting to huge functions - to "fix", in the example, $1s$), contrary [see below] to the standard methods, which, in truncated spaces approach points that ***must*** be veered away form the exact $\psi_n$. This issue is very important, and lacking from standard literature, unless huge functions are successfully used (: infinite complete space ↔ exact eigenfunctions), impractical for large systems.

Furthermore, in avoided crossings of energy surfaces, the functional $F_n$ *recognizes* [5] the flipped root, necessarily appearing within one of the most accurate standard methods, multi-configuration self-consistent field (MCSCF) [7,8], *as the root with the smallest $F_n[root,\{\phi_{i<n}\}]$*, rendering unnecessary to resort to state averaging [9], which, as shown in [5] and demonstrated below, for truncated functions is inaccurate (*in principle*: incorrect, regardless of whether huge wave functions may "fix" it, because the exact state is not *at* the *parameter* crossing [5]). Therefore, as a demonstration, $F_n$ will be used, within Hylleraas coordinates, to compute correct excited state wave functions for He $1s2s$ $^1S$, without needing large expansions of functions, i.e. in convenient small truncated spaces, that would be suitable for large systems.

For completeness of this presentation, proofs for both above issues [3] are indicated in Fig.1. Also the method of identifying a flipped root by $F_n$ within (MCSCF) is demonstrated.

Fig. 1 displays: (i) Three lowest exact (unknown) states mutually orthogonal, $\psi_0,\psi_1,\psi_2$, with energies $E_0<E_1<E_2$. (ii) One (known) approximant, "$\varphi_0$", of the ground state $\psi_0$, and all (trial) functions "$\Phi$" orthogonal to $\varphi_0$ (on the AΦB cycle orthogonal to $\varphi_0$). (iii) The (unknown) function $\varphi_1^+$: the closest to the exact $\psi_1$ among all $\Phi$'s on the AΦB cycle orthogonal to $\varphi_0$, i.e. $\varphi_1^+$ is the Gramm-Schmidt orthogonal to $\varphi_0$ in the subspace of $\{\varphi_0, \psi_1\}$[3]. (All functions are normalized). Evidently [3], $\varphi_1^+$ lies *below* $\psi_1$:

$$\phi_1^+ \equiv (\psi_1 - \phi_0 \langle\psi_1|\phi_0\rangle)/\sqrt{1-\langle\psi_1|\phi_0\rangle^2} \Rightarrow E[\phi_1^+] = E[\psi_1] - (E[\psi_1]-E[\phi_0])\langle\psi_1|\phi_0\rangle^2/(1-\langle\psi_1|\phi_0\rangle^2) \Rightarrow E[\phi_1^+] < E[\psi_1]$$

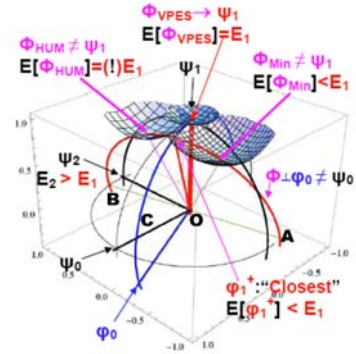

**FIGURE 1.** Schematic representation of normalized states: $\Phi_{HUM}$ ($E[\Phi_{HUM}] = E_1$) and $\Phi_{Min}$ ($E[\Phi_{Min}]<E_1$) are orthogonal to the approximant $\varphi_0$ and are veered away from the exact $\psi_1$. $\Phi_{VPES} \to \psi_1$, independently of the orthogonality to $\varphi_0$ (generally to lower lying approximants $\varphi_i$), and regardless of the accuracy of the latter, i.e. of their closeness to the exact, provided that, if used in VPES $F_n[\varphi_n,\{\varphi_{i<n}\}]$, they are reasonable approximants, as explained in the text [3].

and is *not* a Hamiltonian eigenfunction on the subspace of the AΦB cycle orthogonal to $\varphi_0$, while many other $\Phi$'s on the AΦB cycle orthogonal to $\varphi_0$ lie *above* $\psi_1$, $E[\Phi]>E_1$. Therefore, in going on AΦB cycle, i.e orthogonally to $\varphi_0$, from lower $E[\Phi]<E_1$ to higher $E[\Phi]>E_1$, at least one $\Phi$ has $E[\Phi]=E_1$ (labeled by "$\Phi_{HUM}$" in Fig. 1 - the best, according to HUM theorem, converged 2$^{nd}$ HUM root) having $E[\Phi]=E[\Phi_{HUM}]=E_1$. Actually for truncated expansion approximants, for every $\varphi_0$, the algebraic system of equations $\{\langle\varphi_0|\Phi\rangle = 0$ (orthogonal), $\langle\varphi_0|H|\Phi\rangle = E_1\}$ has many solutions -expansion coefficients of $\Phi$- and none equals $\psi_1$ [3]. $\Phi_{HUM}$, for this $\varphi_0$, is on the AΦB cycle orthogonal to $\varphi_0$, if we consider as $\varphi_0$ the lowest HUM root (which, even worse, in optimizing $\Phi$ as 2$^{nd}$ HUM root, *deteriorates*, compared to any previously independently *optimized lowest root*) [8], therefore the best $\Phi_{HUM}$ is not $\psi_1$.

Besides, we can always diagonalize the Hamiltonian in the AΦB cycle, i.e. in the subspace of $\{\Phi_{HUM}, \varphi_1^+\}$ and this will open their "gap" yielding a lower eigenfunction (labeled $\Phi_{MIN}$ in Fig. 1) lying even lower than $\varphi_1^+$ ($E[\Phi_{MIN}] \leq E[\varphi_1^+]$), on the AΦB cycle orthogonal to $\varphi_0$ [3]. $\Phi_{MIN}$ should be found by a straightforward energy minimization "orthogonally to $\varphi_0$". Therefore, such an energy minimization in a subspace orthogonal to $\varphi_0$ is expected to yield a function lying *lower* than the exact $\psi_1$ and veered away from $\psi_1$. [3] Thus, both "standard" kinds of methods, "based on HUM theorem" and "orthogonal minimization" yield "final" functions which are *veered away form* $\psi_1$ (even from $\varphi_1^+$). [3]

The presented "variational principle" for truncated excited wave functions, $F_n$, are proven [3] to have minimum *at* the exact eigenfunctions (labeled $\Phi_{VPES}$ in Fig. 1) for *any reasonable* lower lying truncated approximants. To the author's knowledge, there is no other method, in the literature, to unambiguously *approach* the exact excited state wave functions in truncated spaces, without resorting to huge wave functions impractical for large systems.

## DEMONSTRATION OF CONVERGING TO THE EXACT $\Psi_1$

The above are demonstrated for He $^1S$ ($1s^2$ and $1s2s$), using Hylleraas variables $\{s=r_1+r_2, t=r_1-r_2, u=|\vec{r}_1-\vec{r}_2|\}$ $u=|\vec{r}_1-\vec{r}_2|$ [1]: $\sum_{i_s,i_t,i_u=0}^{n_s,n_t,n_u} c_{i_s,i_t,i_u} s^{i_s} t^{2i_t} u^{i_u}$, by establishing, a reliable basis $\{\psi_0, \psi_1\} \rightarrow$ the exact $\{\Psi_0, \Psi_1\}$ out of variationally optimized state-specific Laguerre-type orbitals [2]: $\sum_{k=0}^{n-1} a_{n,k} \left(-2rz_n/n\right)^k z_n^{3/2} e^{-rz_n/n}$ (via $s \pm t$) whose polynomial coefficients and exponents are optimized, in a Slater determinant, thus achieving accurate (27 term - compared to Pekeris' 95 terms [10]) series expansions $\{\psi_0, \psi_1\}$, in terms of $s^i t^{2j} u^k$, ($E_0 = -2.90371$, $E_1 = -2.14584$ a.u.), in order to compare (project on it) the demonstrated truncated approximants of up to 8 terms ($s^1 t^2 u^1$).

**TABLE 1**. Optimized 8-term truncated approximants, $\Phi$, (up to: $s^1 t^2 u^1$)

| $\Phi$ | $E_0$ (a.u.) | $E_1$ (a.u.) | $\langle\psi_0|\Phi\rangle$ | $\langle\psi_1|\Phi\rangle$ |
|---|---|---|---|---|
| $\psi^{(*)}$ | -2.90371 | -2.14584 | | |
| $\Phi_0^{1r\ HUM}$ | -2.90312 | (E[$^{2r}$] -2.02)$^{(1)}$ | 0.99996 | 0.00001 |
| $\Phi_1^{2r\ HUM}$ | (E[$^{1r}$] -2.897)$^{(1)}$ | **-2.14449** | 0.0033 | 0.9986 |
| $\Phi_1^{F[\varphi o]}$ | E[$\varphi_o$] -2.848$^{(2)}$ | **-2.14515** | 0.0049 | **0.9998** |
| $\Phi_1^{'F[\varphi'o]}$ | E[$\varphi'_o$] -2.842$^{(2)}$ | **-2.1452** | | |

$^{(*)}$ Optimized 27-term (up to: $s^2 t^4 u^2$)
$^{(1)}$ Other eigenvalues of the secular matrix concomitant to the optimized (deteriorated)
$^{(2)}$ 1-term $\varphi_o$s (up to: $s^0 t^0 u^0$) used in $F_1$ (i.e. $F_n$, n=1, in Eq.1)

Beyond the above established $\{\psi_0, \psi_1\}$, Table 1 shows the optimized 8-term approximants and their overlaps with $\{\psi_0, \psi_1\}$ of: (i) The ground state $\Phi_0^{1r\ HUM}$ along with its orthogonal deteriorated 2$^{nd}$ root. (ii) The 2$^{nd}$ HUM root $\Phi_1^{2r\ HUM}$. Its main two orbitals **resemble 1s1s'** (rather than 1s2s); its orthogonal deteriorated 1$^{st}$ root is also shown. (iii) The functions $\Phi_1^F$ obtained by *minimizing* $F_1$, by using, in $F_1$, two different, rather inaccurate, 1-term lower approximants $\varphi_0$ (also shown). Contrary to the aforesaid 2$^{nd}$ HUM root $\Phi_1^{2r\ HUM}$, now the main orbitals **resemble 1s2s**, as expected, and, as seen by the overlaps, $\Phi_1^F$ is much closer to $\psi_1$ than $\Phi_1^{2r\ HUM}$. Thus, the truncated approximant $\Phi_1^F$, obtained by *minimizing* $F_1$, reliably approaches $\psi_1$.

## DEMONSTRATION OF IDENTIFYING A "FLIPPED ROOT"

Parametrize the ground and 1$^{st}$ excited wave functions of a hydrogen-like ion as

$$\psi_0(z_0;r) = a_0(z_0) e^{-Z r z_0}, \quad \psi_1(z_1,g;r) = a_1(z_1,g) e^{-z_1 Z r/2}(1 - g Z r/2) \quad (2)$$

where $Z$ is the nuclear charge, ($z_0, z_1, g$) are variational parameters and $a_0(z_0)$, $a_1(z_1,g)$ are normalization constants. These functions are not orthonormal, unless $z_0 = 1$, $z_1 = 1$, $g = 1$, when they form eigenfunctions of the Hamiltonian $H\psi_j(r) = -\psi_j''(r)/2 - \psi_j'(r)/r - Z\psi_j(r)/r$. In their 2 x 2 subspace create an orthonormal basis whose overlap matrix is $\delta_{i,j}$: $\{\Psi_0(r) = \psi_0(r), \Psi_1(r) = (\psi_1(r) - \psi_0(r)\langle\psi_0|\psi_1\rangle)/\sqrt{1-\langle\psi_0|\psi_1\rangle^2}\}$. Let $\{\Phi^{1r}, \Phi^{2r}\}$ be the two normalized eigenfunctions of the Hamiltonian matrix $\langle\Psi_i|H|\Psi_j\rangle = \int_0^\infty 4\pi r^2 \Psi_i(r) H \Psi_j(r) dr$, with their eigenvalues ("roots") depending on ($z_0, z_1, g$).

Now, around $z_0 \approx 2$, a root crossing occurs for a wide range around $z_1 \approx 1$, and $g \approx 1$. Near and "before" the crossing the continuation of $\psi_0(z_0,r)$ is $\Phi^{1r}(r)$, and the continuation of $\psi_1(z_1,g;r)$ is $\Phi^{2r}(r)$, whereas "beyond" the crossing the lowest $\Phi^{1r}(r)$ is the continuation of $\psi_1(z_1,g;r)$ and the continuation of $\psi_0(z_0,r)$ is $\Phi^{2r}(r)$. The question is to decide, via $F_1$, whether a given value of ($z_0, z_1, g$), near the crossing, is "before" or "beyond" the crossing, in order to use the continuation, $\varepsilon$, of (always) $E[\psi_1(z_1,g;r)]$ in an optimization algorithm. (In the present demonstration by Newton-Raphson (NR), $\boldsymbol{\varepsilon}' = \nabla_{(z_0,z_1,g)}\varepsilon = 0$ is solved by proceeding iteratively to a new point $\mathbf{p} + \delta\mathbf{p} = \mathbf{p} - \mathbf{J}^{-1} \cdot \boldsymbol{\varepsilon}'$ -or less if the method diverges-having started at some point $\mathbf{p}=(z_0, z_1, g)$, where $\mathbf{J}$ is the Jacobian matrix - Hessian of $\varepsilon$.)

Thus, consider $F_1$ (i.e. $F_n$, n=1, in Eq.1) for both "roots", $\Phi^{1r}$, $\Phi^{2r}$, and use a fixed predetermined (deliberately not very accurate) ground state approximant $\varphi_0$ with $z_0 = 1-0.05$. ($F_1[\varphi_0; \Phi^{2r}(r)]$ is directly minimized at $z_1 = 1$, and $g = 1$, ($z_1 = 0.95 \approx 1$), to $F_1 \rightarrow E[\Phi^{2r}(r)] = E[\psi_1(1,1;r)] = -0.125$.) Among the two "roots", *the continuation of the*

*excited state, near the crossing, is the one with the lowest $F_1$*. Indeed, for $Z=1$, using (first trial traditionally:) always the "2$^{nd}$ root", i.e. keeping $\varepsilon$ to $E^{2r}=E[\Phi^{2r}(r)]$, [regardless of which $n$ ($n^{th}$ root) the lowest $F_1$ suggests] "root-flipping" shows up: Even by using half NR-step: p=(2.7, 0.95, 0.8), n=1, $E^{2r}$= 1.09915 is obtained, the sequence of p-values *does not converge*, whereas, by consulting $F_1$ the continuation of the excited state is recognized near the crossing and used (until finally, at convergence, only $n=2$, the 2$^{nd}$ root, is suggested by the *lowest $F_1$*) cf. case (i) of Table 2. Observe that at the beginning, "beyond" the crossing, the lowest $F_1$ dictates to use, for the next step, the (*lower* than $E_1$) value of $\varepsilon = E[\Phi^{1r}(r)]$=-0.1423 ($n=1$, the **lowest** wave function at that point).

TABLE 2. The lowest $F_1$ recognizes which n$^{th}$ root approximates the excited state near a crossing

| (i) E$^{2r}$ **fails**, $F_1$ succeeds | | | (ii) E$^{2r}$ succeeds accidentally-$F_1$ ignored | | | (iii) $F_1$ succeeds (same starting **p**) | | |
|---|---|---|---|---|---|---|---|---|
| **p** | n | ε | **p** | n | E$^{2r}$ | **p** | n | ε |
| (2.70, 0.95, 0.80) | 1 | **-0.1423** | (2.70, 1.20, 1.10) | 1 | 0.9948 | (2.70, 1.20, 1.10) | 1 | -0.1197 |
| (2.63, 0.96, 0.91) | 1 | **-0.1272** | ... | | | (1.94, 1.03, 0.98) | 1 | **-0.1262** |
| ... | | | (1.59, 0.52, 0.72) | 2 | -0.0866 | (1.77, 0.89, 1.08) | 2$^{(+)}$ | -0.1213 |
| (1.93, 0.99, 1.01) | 1 | **-0.1251** | (-3.06, -10.9, -12.3) | 1 | 69.042 | ... | | ... |
| (1.32, 1.03, 0.95) | 2$^{(+)}$ | -0.1249 | ... | | | (1.04, 0.96, 1.19) | 2 | -0.1249 |
| ... | | ... | (-0.01, -0.03, 0.02) | 2 | 0.0214 | ... | | ... |
| (0.91, 1.00, 0.98) | 2 | -0.1250 | ... | | | (0.74, 0.99, 1.08) | 2 | -0.1250 |
| (0.84, 1.00, 0.98) | 2 | **-0.125** | (0.52, 1.00, -0.43) | 2 | **-0.125** | (0.70, 1.00, 1.07) | 2 | **-0.125** |
| $^{(+)}$ Root recognized, lowest $F_1$ suggests: For the next step "take" n=2, the 2$^{nd}$ root | | | | | | | | |

Similarly, using only the 2$^{nd}$ root (not dictated by the lowest $F_1$) and starting, again "beyond" the crossing ($n=1$) [cf. case (ii) in Table 2], despite the original irregularities due to root-flipping, the 2$^{nd}$ root finally happened to remain "before" the crossing ($n=2$), and converged; 0.3 of NR-step was used. Now, by consulting $F_1$ no irregularities occurred [cf. case (iii) in Table 2]. Note that finally, near the minimum of the 2$^{nd}$ root, where $E[\Psi_1(\mathbf{p}_1)]>E[\Psi_1(\mathbf{p}_1)]$ "before" the crossing, the convergence *should* use the 2$^{nd}$ root.

Note that the graphs (Eq.2) of all converged functions, above, are practically identical to the exact.

## ACKNOWLEDGMENTS

Financial support of this work by the General Secretariat for Research and Technology, Greece, (project Polynano-Kripis 447963) is gratefully acknowledged.

## REFERENCES

1. E. Hylleraas and B. Undheim, *Z. Phys.* **65,** 759; (1930) J. K. L. McDonald, *Phys. Rev.* **43**, 830 (1933).
2. Z. Xiong and N. C. Bacalis, *Commun. Math. Comput. Chem.* **53**, 283 (2005); Z. Xiong, M. Velgakis, and N. C. Bacalis, *Int. J. Q. Chem.*, **104**, 418 (2005).
3. N. C. Bacalis, "Utilizing the fact ..." in *Computational Methods in Science and Engineering-2007*, edited by T. E. Simos et al., AIP Conference Proceedings 963, American Institute of Physics, Melville, NY, 2007, vol. 2, part A, pp. 6-9; N. C. Bacalis, "Remarks on the Hylleraas-Undheim and MacDonald higher roots, and functionals having local minimum at the excited states" in *Computational Methods in Science and Engineering-2008*, edited by T. E. Simos et al., AIP Conference Proceedings 1148, American Institute of Physics, Melville, NY, 2009, Vol. 2, pp. 372-375; N. C. Bacalis, Z. Xiong and D. Karaoulanis, *J. Comput. Meth. Sci. Eng.*, **8**, 277 (2008).
4. Z. Xiong and N.C. Bacalis, *Chinese Physics B*, **19**, 023601 (2010); Z. Xiong, Z.-X. Wang and N. C. Bacalis, *Acta Phys. Sin.*, **63**, 053104 (2014); unpublished (*Int. J. Q. Chem.*)
5. N. C. Bacalis, unpublished (ICCMSE-2015).
6. H. Shull and P. -O. Löwdin, Phys. Rev. 110, 1466 (1958)
7. P. -O. Löwdin, *Phys. Rev.*, **97**, 1509 (1955); P. -O. Löwdin, *Adv. Chem. Phys.*, **2**, 207 (1959);
8. R. Shepard, *Adv. Chem. Phys.*, **69**, 63 (1987); H. -J. Werner, *Adv. Chem. Phys.* **69**, 1 (1987); H. -J. Werner, W. Meyer, *J. Chem. Phys.*, **73**, 342 (**1980**); M. Frisch, I. Ragazos, M. Robb, H. Schlegel, *Chem. Phys. Lett.* **189**, 524 (1992);
9. L. Cheung, S. Elbert, K. Ruedenberg, *Int. J. Quantum Chem.*, **16**, 1069 (1979); K. Docken, J. Hinze, *J. Chem. Phys.*, **57**, 4928 (1972); H. -J. Werner, W. Meyer, *J. Chem. Phys.*, **74**, 5794 (1981); D. Yeager, D. Lynch, J. Nichols, P. Jørgensen, J. Olsen, *J. Phys. Chem.*, **86**, 2140 (1982); J. Olsen, P. Jørgensen, D. Yeager, *J. Chem. Phys.*, **76**, 527 (**1982**); J. Golab, D. Yeager, P. Jørgensen, *Chem. Phys.* **1985**, 93, 83.
10. C. L. Pekeris, *Phys. Rev.* **126**, 1470 (1962).